\documentclass[preprint,showpacs,amsmath,amssymb]{revtex4}

\usepackage{graphics}
\usepackage{dcolumn}
\usepackage{bm}

\begin{document}

\title{Statistical mechanics of a colloidal suspension in contact with a
fluctuating membrane}

\author{T. Bickel}
\altaffiliation[Corresponding author. Electronic address: ]{th.bickel@cpmoh.u-bordeaux1.fr}
\affiliation{Centre de Physique Mol\'{e}culaire Optique et Hertzienne, Universit\'{e} Bordeaux 1, 
UMR 5798, 351 cours de la Lib\'eration, 33405 Talence, France}

\author{M. Benhamou}
\affiliation{Laboratoire de Physique des Polym\`{e}res et Ph\'{e}nom\`{e}nes Critiques, \\
Facult\'{e} des Sciences Ben M'sik, B.P. 7955, Casablanca, Morocco }

\author{H. Ka\"{\i}di}
\affiliation{Laboratoire de Physique des Polym\`{e}res et Ph\'{e}nom\`{e}nes Critiques, \\
Facult\'{e} des Sciences Ben M'sik, B.P. 7955, Casablanca, Morocco }

\date{\today}

\begin{abstract}
Surface effects are generally prevailing in confined colloidal systems. Here
we report on dispersed nanoparticles close to a fluid membrane. Exact
results regarding the static organization are derived for a dilute solution
of non-adhesive colloids. It is shown that thermal fluctuations of the
membrane broaden the density profile, but on average colloids are neither
accumulated nor depleted near the surface. The radial correlation function
is also evaluated, from which we obtain the effective pair-potential between
colloids. This entropically-driven interaction shares many similarities with
the familiar depletion interaction. It is shown to be always attractive with
range controlled by the membrane correlation length. The depth of the
potential well is comparable to the thermal energy, but depends only
indirectly upon membrane rigidity. Consequenses for stability of the
suspension are also discussed.
\end{abstract}


\maketitle


\section{Introduction}

\label{sec:intro}

Fluid membranes are soft surfaces, self-assembled from surfactant solutions~ 
\cite{safranbook}. They assume a large variety of shapes and topologies,
which have been accurately explained in terms of bending energy~\cite
{helfrich73,seifert97}. In most practical realizations, however, membrane
suspensions are not pure but incorporate colloidal entities as well. In
living systems for instance, lipid bilayers organize the cell into
compartments that keep apart different chemical environments~\cite
{lodishbook}. Biological membranes are therefore in contact with various
kinds of proteins, macro-ions or more complex structures~\cite{fradinBPJ03}.
Membrane phases used in detergency or cosmetics also include numerous
additives, like macromolecules or colloids, in order to improve efficiency
and to control viscoelastic properties~\cite{wennerbook}. One is then
naturally inclined to investigate transformations that may occur upon
addition of colloids, and several studies have recently been devoted to
these complex systems. Essentially, these have focused on the softening of
membrane resulting from the depletion of spherical and rod-like colloids~ 
\cite{yamanPRL97}, or the depletion~\cite{podgornikEPL93,hankePRE99} and
adsorption~\cite{brooksJPII91,clementJPII97,skauEPJE02} of flexible
polymers, but the predicted effects are usually small compared to the bare
rigidity of the bilayer. Bending of a membrane upon colloid adsorption has also
drawn growing interest because of potential applications for drug
encapsulation and gene delivery~\cite
{radlerScience97,dietrichJPII97,koltoverPRL99,lipowskyEPL98,desernoEPL03}.
Nevertheless, a common feature shared by most theoretical studies is that
membrane fluctuations are systematically disregarded. This point clearly
illustrates the technical difficulty to couple bulk and surface degrees of
freedom.

Generally, the mutual influence of bulk and surface properties on each other
is a challenging problem~\cite{louisJCP02,netzPR03}. The situation has now
been clarified for bidispersed hard-sphere suspensions~\cite
{bibenJPCM96,rothPRE00,likosPR01}: when in contact with a flat substrate,
excluded-volume effects are known to push the larger beads toward the wall
of the sample~\cite{kaplanPRL94}. Recent experiments done with curved or
corrugated surfaces have shown that geometric features of the surface can
also create and modulate entropic force fields~\cite
{dinsmoreNature96,dinsmorePRL98}. These depletion forces can be used to grow
oriented colloidal crystal, with numerous potential applications such as the
fabrication of photonic bandgap crystals~\cite{linPRL00}. The theoretical
description of those systems usually requires advanced density functional
techniques~\cite{rothPRE00}, that have been adapted to study depletion
potentials close to arbitrarily shaped substrates~\cite{brykEPL03}. In this
work, we follow a different line and present some new findings regarding the
static organization of nanoparticles near a \emph{fluctuating} surface.
Given the increasing complexity of the problem, we focus on the simplest
system consisting of a monodispersed, dilute solution of non-adhesive
colloids. This allows us to derive exact results and to highlight non-trivial
phenomena, such as membrane-induced interaction between the colloids.

This paper is organized as follows. In Sec.~\ref{sec:partition}, we present
an exact computation of the partition function of the global system. The
determination of the variation of the particle density upon distance is the
aim of Sec.~\ref{sec:profile}. The computations of the radial distribution
function and the resulting pair-potential and related discussion are
presented in Sec.~\ref{sec:pair}. We draw some concluding remarks in the
last section. Finally, some technical details are relegated to the
appendices.

\section{The partition function}

\label{sec:partition}

Consider a fluctuating membrane in contact with a colloidal suspension,
consisting of nanoparticles immersed in some solvent. We assume that the
colloids cannot permeate through the bilayer on experimental time-scales, in
such way that the membrane act as a flexible but impenetrable wall for the
particles. In this paper, we shall use the notation $\mathbf{r}=\left( {%
\boldsymbol{\rho} },z\right) $, where ${\boldsymbol{\rho} }=(x,y)$ is the
transverse vector and $z$ the perpendicular component. The position of the
(almost flat) membrane is specified through the displacement field $h\left(
x,y\right) $. The surface fluctuates around the horizontal plane $z=0$, so
that the height $h\left( x,y\right) $ may take either positive or negative
values. The equilibrium statistical mechanics of membranes is based on the
Helfrich Hamiltonian~\cite{helfrich73} 
\begin{equation}
\mathcal{H}_0\left[ h\right] =\frac{1}{2}\int d^{2}{\boldsymbol{\rho} }\left[
\kappa \left( \Delta h\right) ^{2}+\sigma \left( \nabla h\right) ^{2}+\mu
h^{2}\right] \ ,  \label{helfrich}
\end{equation}
where $\left( \kappa ,\sigma ,\mu \right) $ are the elastic constants of the
membrane. In what follows, we shall rather use the rescaling parameters: $%
\widehat{\kappa }=\beta \kappa $, $\widehat{\sigma }=\beta \sigma $, and $%
\widehat{\mu }=\beta \mu $, where we define as usually $\beta =1/k_{B}T$,
with $T$ the absolute temperature and $k_{B}$ the Boltzmann constant. Let us
also introduce the height-height correlation function 
\begin{equation}
G\left( {\boldsymbol{\rho} }-{\boldsymbol{\rho} }^{\prime }\right)
=\left\langle h\left( {\boldsymbol{\rho} }\right) h \left( {\boldsymbol{\rho}
}^{\prime }\right) \right\rangle _{0}-\left\langle h\left( {\boldsymbol{\rho}
}\right) \right\rangle _{0}\left\langle h\left( {\boldsymbol{\rho} }^{\prime
}\right) \right\rangle _{0} \ ,  \label{defpropagator}
\end{equation}
from which one obtains the mean-squared fluctuations $\xi _{\perp
}^2=G\left( 0\right)$. Some properties of $G$ are recalled in Appendix~%
\ref{app:a}.

We intend to compute some statistical properties of the system under
investigation, namely the particle density profile and radial distribution
function~\cite{hansenbook}. For a dilute solution, one may ignore the
interactions between particles and treat them as an ideal gas in confined
geometry. Since we are interested in those systems containing a considerable
number of particles (thermodynamic limit), the physical quantities are
independent of the particular choice of statistical ensemble. For
convenience, we shall consider both the \emph{canonical} and the \emph{grand
canonical} ensembles. The grand canonical partition function, denoted $ \mathcal{Z}_{G}$%
, is the Laplace transform of the canonical partition function, that is 
\begin{equation}
\mathcal{Z}_{G}=\sum_{N=0}^{+\infty }f^{N}\mathcal{Z}_{c}\left( N\right) \ ,  \label{defgrandcan}
\end{equation}
with $f$ the fugacity. In the above equality, the canonical partition
function $\mathcal{Z}_{c}\left( N\right) $ is 
\begin{equation}
\mathcal{Z}_{c}\left( N\right) =\frac{1}{\lambda ^{3N}N!}\int d^{3}\mathbf{r}%
_{1}...d^{3} \mathbf{r}_{N}\int \mathcal{D}he^{-\beta \mathcal{H}\left[ h%
\right] }e^{-\beta \sum_{i=1}^{N}V\left( \mathbf{r}_{i}\right) } \ .
\label{defcan}
\end{equation}
The functional integral extends over all configurations of the field $h$,
weighted with the Helfrich energy~(\ref{helfrich}). The positions $\mathbf{R}%
_{N}=\left( \mathbf{r}_{1},...,\mathbf{r}_{N}\right) $ of the $N$ particles
are restricted to the upper side of the three-dimensional space limited by
the membrane. In the above equation, the thermal wavelength $\lambda $
results from integration over the particles momenta, and colloid-membrane
interactions are accounted for through the (contact) hard core potential $%
V\left( \mathbf{r}\right) $ 
\begin{equation}
V\left( \mathbf{r}_{i}\right) =\left\{ 
\begin{array}{lll}
0\text{ }, &  & \text{if }z_{i}>h\left( x_{i},y_{i}\right) \ , \\ 
+\infty \ , &  & \text{otherwise }.
\end{array}
\right.  \label{hardcore}
\end{equation}

With these considerations, we first examine the canonical partition
function. It can be written 
\begin{equation}
\label{grandcan0}
\mathcal{Z}_{c}=\frac{\mathcal{Z}_{0}}{\lambda ^{3N}N!}\int \prod_{i=1}^{N}d^{3}\mathbf{r}%
_{i}\varphi _{N}\left( \mathbf{R}_{N}\right) \ ,
\end{equation}
where it is convenient to define the function $\varphi _{N}$ 
\begin{equation}
\varphi _{N}\left( \mathbf{R}_{N}\right) =\mathcal{Z}_{0}^{-1}\times \int \mathcal{D}%
he^{-\beta \mathcal{H}\left[ h\right] } \prod_{i=1}^{N}\theta \left[
z_{i}-h\left( x_{i},y_{i}\right) \right] \ .  \label{phindef}
\end{equation}
Here, $\mathcal{Z}_{0}=\int \mathcal{D}h\exp \left\{ -\beta \mathcal{H}\left[ h\right]
\right\} $ is the partition function of a membrane in the absence of
particles ($N=0$), and $\theta \left( x\right) $ is the step function. It is
easy to see, from its definition~(\ref{phindef}), that $\varphi _{N}$
satisfies the following boundary conditions at infinity 
\begin{eqnarray}
\left. \varphi _{N}\left( {\boldsymbol{\rho} }_{1},...,{\boldsymbol{\rho} }%
_{N};z_{1},...,z_{N}\right) \right| _{z_{1}=...=z_{N}=-\infty }=0 \ ,
\label{bcminus} \\
\left. \varphi _{N}\left( {\boldsymbol{\rho} }_{1},...,{\boldsymbol{\rho} }%
_{N};z_{1},...,z_{N}\right) \right| _{z_{1}=...=z_{N}=+\infty }=1 \ ,
\label{bcplus}
\end{eqnarray}
for fixed values of the transverse vectors $\left( {\boldsymbol{\rho} }%
_{1},...,{\boldsymbol{\rho} }_{N}\right) $. The function $\varphi _{N}\left( 
\mathbf{R}_{N}\right) $ has to be understood as the partition function of a
membrane whose configurations are subjected to $N$ restrictions $h\left(
x_{i},y_{i}\right) \leq z_{i}$, with $1\leq i\leq N$. It is generally a
complicated function of the relative transverse distances $\left| {%
\boldsymbol{\rho} }_{i}-{\boldsymbol{\rho} }_{j}\right| $ and perpendicular
components $z_{i}$. Indeed, the translation symmetry is preserved in the
parallel directions, but due to the presence of the membrane, this symmetry
is broken in the perpendicular one. After some algebra detailed in Appendix~%
\ref{app:b}, we find that $\varphi _{N}$ is given by 
\begin{equation}
\varphi _{N}\left( \mathbf{R}_{N}\right) =\left( 2\pi \right) ^{-N/2}\left[
\det \mathcal{G}_N\right] ^{-1/2}\int_{-\infty }^{z_{1}}dz_{1}^{\prime
}...\int_{-\infty }^{z_{N}}dz_{N}^{\prime }\exp \left\{ -\frac{1}{2}
\sum_{i,j=1}^{N}z_{i}^{\prime }.\left[ \mathcal{G}_N^{-1}\right]
_{ij}.z_{j}^{\prime }\right\} \ ,  \label{phin}
\end{equation}
where $\mathcal{G}_N$ is the squared matrix of order $N$ whose elements are
the propagators $[\mathcal{G}_N]_{ij}=G\left( {\boldsymbol{\rho} }_{i}-{%
\boldsymbol{\rho} }_{j}\right) $ defined in Eq.~(\ref{defpropagator}). This
expression is compatible with the boundary conditions (\ref{bcminus}) and (%
\ref{bcplus}), and more generally we can deduce from Eq.~(\ref{phin}) the
fundamental property according to which 
\begin{equation}
0\leq \varphi _{N}\left( {\boldsymbol{\rho} }_{1},...,{\boldsymbol{\rho} }%
_{N};z_{1},...,z_{N} \right) \leq 1\text{ },
\end{equation}
for all values of the position vectors $\left( \mathbf{r}_{1},...,\mathbf{r}%
_{N}\right) $. The right-hand side inequality relies on the fact that $%
\varphi _{N}$ is an \emph{incomplete} Gaussian multiple-integral.

Now, we direct our attention to the grand canonical function. Formally, it
writes as 
\begin{eqnarray}
\mathcal{Z}_{G} &=&\sum_{N=0}^{+\infty }f^{N}\frac{1}{\lambda ^{3N}N!}\int \mathcal{D}%
h e^{-\beta \mathcal{H}\left[ h\right] }\int \prod_{i=1}^{N}d^{3}\mathbf{r}%
_{i}\theta \left[ z_{i}-h\left( {\boldsymbol{\rho} }_{i}\right) \right] 
\nonumber \\
&=&\int \mathcal{D}he^{-\beta \mathcal{H}\left[ h\right] }\sum_{N=0}^{+%
\infty }f^{N}\frac{1}{\lambda ^{3N}N!}\left\{ \int d^{3}\mathbf{r}\theta %
\left[ z-h\left( {\boldsymbol{\rho} }\right) \right] \right\} ^{N}  \nonumber
\\
&=&\int \mathcal{D}he^{-\beta \mathcal{H}\left[ h\right] }\exp \left\{ \frac{%
f}{\lambda ^{3}}\int d^{3}\mathbf{r}\theta \left[ z-h\left( {%
\boldsymbol{\rho} }\right)\right] \right\} \ .
\end{eqnarray}
Integrating over the $z-$variable then yields 
\begin{equation}
\mathcal{Z}_{G}=\mathcal{Z}_{0}e^{f\Omega /2\lambda ^{3}}\times \frac{\int \mathcal{D}he^{-\beta 
\mathcal{H}\left[ h\right] -f\lambda ^{-3}\int d^{2}{\boldsymbol{\rho} }%
h\left( {\boldsymbol{\rho} }\right) }}{Z_{0}} \ .  \label{pression}
\end{equation}
Here, $\Omega =S\times L$ is the total volume occupied by the system under
investigation, $S$ is the area of the horizontal plane and $L/2$ represents
the upper bound of the perpendicular coordinate $z$. The above functional
integral can be easily calculated, and we find 
\begin{equation}
\frac{\int \mathcal{D}he^{-\beta \mathcal{H}\left[ h\right] -f\lambda
^{-3}\int d^{2}{\boldsymbol{\rho} }h\left( {\boldsymbol{\rho} }\right) }}{%
\mathcal{Z}_{0}}=\exp \left\{ f^{2}S \widetilde{G}\left( 0\right) /2\lambda
^{6}\right\} \ ,
\end{equation}
with $\widetilde{G}\left( 0\right) =1/\widehat{\mu }$ the Fourier transform
of the propagator $G$ at $q=0$. To obtain this result, we have applied the
general formula~(\ref{result}) of Appendix $B$ to the particular source $%
J\left( {\boldsymbol{\rho} }\right) =-f\lambda ^{-3}$. Finally, we have the
simple, \emph{exact} expression for the partition function $\mathcal{Z}_{G}$ 
\begin{equation}
\mathcal{Z}_{G}=\mathcal{Z}_{0}\times \exp \left\{ \frac{f\Omega }{2\lambda ^{3}}\right\} \times
\exp \left\{ \frac{f^{2}S}{2\widehat{\mu }\lambda ^{6}}\right\} \ .
\label{grandcan}
\end{equation}
Noticeably, $\mathcal{Z}_{G}$ splits into three parts. The first two factors are the
standard contributions: $\mathcal{Z}_{0}$ is the partition function of the membrane in
a particle-free solvent, and $\exp \left\{ f\Omega /2\lambda ^{3}\right\} $
is the partition function of an ideal gas of colloids (remember that the
particles are restricted to the upper half of the space). All the
information concerning the interplay between bulk and surface contributions
finally factorizes in the last term of Eq.~(\ref{grandcan}).

With this expression of $\mathcal{Z}_{G}$, we are now able to evaluate the average
number of colloids 
\begin{equation}
\left\langle N\right\rangle =f\frac{\partial \ln \mathcal{Z}_{G}}{\partial f}=\frac{
f\Omega }{2\lambda ^{3}}\times \left( 1+\frac{2f}{\widehat{\mu }L\lambda ^{3}%
}\right) \ .  \label{number}
\end{equation}
In addition to the usual term $f\Omega /2\lambda ^{3}$, we find a second
contribution that happens to be negligible at low concentration $\left(
f<<1\right) $ or for very large perpendicular extension $\left( L>>\widehat{%
\mu }^{-1}\lambda ^{-3}\right) $. For the sake of simplicity, we will assume
thereafter that these requirements are fulfilled so that the bulk
concentration is 
\begin{equation}
\rho _{\infty }=\frac{\left\langle N\right\rangle }{\Omega /2}\simeq f\lambda
^{-3} \ .  \label{bulkdensity}
\end{equation}
Note, however, that the second contribution in Eq.~(\ref{number}) could not
be disregarded in a strongly confining system. This point might be relevant
for experimental realizations involving colloids in a lamellar phase~\cite
{bougletEPJB99,imaiEPJE04} or in a sponge phase~\cite{tanakaPRL02} of
membranes.

\section{Particle density profile}

\label{sec:profile}

\begin{figure}[tbp]
\begin{center}
\includegraphics{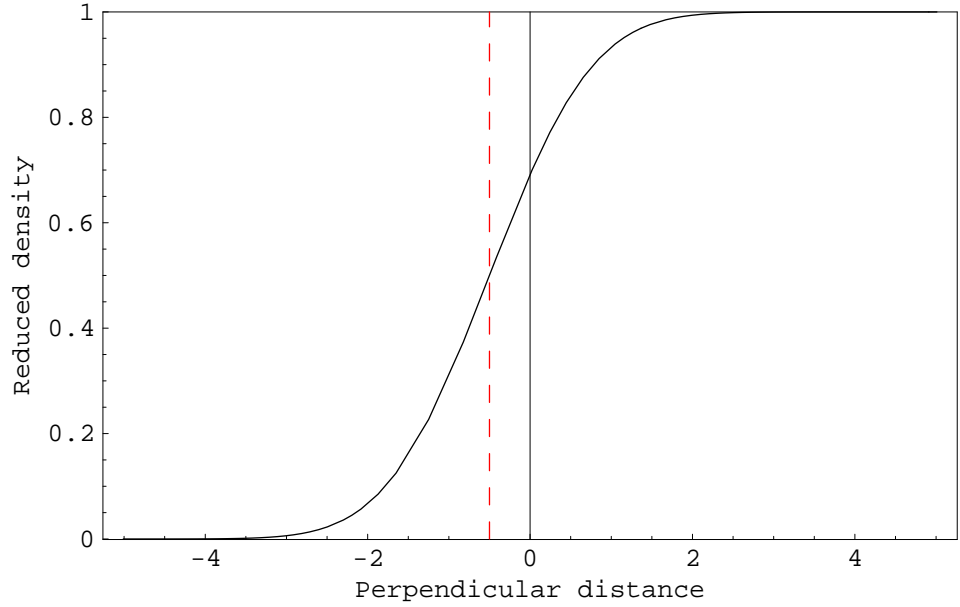}
\end{center}
\par
\caption{Reduced density $\rho (z)/\rho _{\infty }$ as a
function of the reduced perpendicular distance $z/\xi _{\perp }$.
The density profile is symetric with respect of the dotted line $z=-z_{0}$,
so that the number of particles that enter the depressions of the undulating
surface exactly compensate for the depleted ones.}
\label{fig1}
\end{figure}

The concentration profile of an ideal gas of colloids in contact with a
rigid wall located at $z=0$ is simply $\rho _{HW}=\rho _{\infty }\times
\theta (z)$. For a flexible interface, the situation is quite different as
thermal undulations are expected to broaden the distribution. In this
section, we examine the mean-value of the particle density at point $\mathbf{%
r}$ defined by 
\begin{equation}
\rho \left( \mathbf{r}\right) =\left\langle \sum_{i=1}^{N}\delta _{3}\left( 
\mathbf{r}-\mathbf{r}_{i}\right) \right\rangle =N\left\langle \delta
_{3}\left( \mathbf{r}-\mathbf{r}_{1}\right) \right\rangle \ ,
\label{defdensity}
\end{equation}
with $\delta_3$ the three-dimensional Dirac distribution.
The averaging procedure implies integration over the colloid configurations
as well as internal degrees of freedom of the membrane. Explicitly, we have 
\begin{eqnarray}
\rho \left( \mathbf{r}\right) &=&\frac{1}{\mathcal{Z}_{G}}\sum_{N=0}^{+\infty }\frac{%
f^{N}} {\lambda ^{3N}N!}N\int \mathcal{D}he^{-\beta \mathcal{H}\left[ h%
\right] }\int \prod_{i=1}^{N-1}d^{3}\mathbf{r}_{i}\theta \left[
z_{i}-h\left( x_{i},y_{i}\right) \right] \theta \left[ z-h\left( {%
\boldsymbol{\rho} }\right) \right]  \nonumber \\
&=&\frac{f\lambda ^{-3}}{\mathcal{Z}_{G}}\int \mathcal{D}he^{-\beta \mathcal{H}\left[
h \right] }\exp \left\{ \frac{f}{\lambda ^{3}}\int d^{3}\mathbf{r}^{\prime
}\theta \left[ z^{\prime }-h\left( {\boldsymbol{\rho} }^{\prime }\right) %
\right] \right\} \theta \left[ z-h\left( {\boldsymbol{\rho} }\right) \right]
\nonumber \\
&=&\mathcal{Z}_{0}\frac{f\lambda ^{-3}}{\mathcal{Z}_{G}}e^{f\Omega /2\lambda ^{3}}\times \frac{%
\int \mathcal{D}he^{-\beta \mathcal{H}\left[ h\right] -f\lambda ^{-3}\int
d^{2}{\boldsymbol{\rho} }^{\prime }h\left( {\boldsymbol{\rho} }^{\prime
}\right) }\theta \left[ z-h\left( {\boldsymbol{\rho} }\right) \right] }{\mathcal{Z}_{0}%
} \ .
\end{eqnarray}
The last functional integral can be computed making a simple translation of
the field $h\rightarrow h+f/\lambda ^{3}\widehat{\mu }$. We finally get 
\begin{equation}
\rho \left( \mathbf{r}\right) =f\lambda ^{-3}\times \varphi _{1}\left( {%
\boldsymbol{\rho} };z+f/\lambda ^{3}\widehat{\mu }\right) \ ,
\end{equation}
where $\varphi _{1}$ is a particular function of type~(\ref{phin}), that is 
\begin{equation}
\varphi _{1}\left( {\boldsymbol{\rho} };z+f/\lambda ^{3}\widehat{\mu }%
\right) =\left[ 2\pi G\left( 0\right) \right] ^{-1/2}\int_{-\infty
}^{z+f/\lambda ^{3} \widehat{\mu }}dz^{\prime }\exp \left\{ -\frac{1}{2}%
G^{-1}\left( 0\right) z^{2}\right\} \ .
\end{equation}
For $N=1$, the squared matrix $\mathcal{G}_1$ in relation~(\ref{phin})
reduces to $G\left( 0\right) =\xi _{\perp }^{2}$. Recalling that $f\lambda
^{-3}$ equals the bulk density $\rho _{\infty }$, we find the final
expression for the density profile 
\begin{equation}
\rho \left( z\right) =\rho _{\infty }\times \frac{1}{2}\left[ 1+\text{erf}
\left( \frac{z+z_{0}}{\sqrt{2}\xi _{\perp }}\right) \right] \ ,
\label{density}
\end{equation}
with the characteristic length 
\begin{equation}
z_{0}=\rho _{\infty }\widehat{\mu }^{-1} \ .  \label{defz0}
\end{equation}
The concentration profile is shown in fig.~(\ref{fig1}). As expected, it
only depends on the perpendicular distance $z$ (homogeneity property in the
parallel directions). For fixed parameter $\widehat{\mu }$, the scale $z_{0}$
becomes smaller as the particle density is decreased. The physical meaning
of this length can be understood as follows. When in contact with the
colloidal solution, the membrane experiences the osmotic pressure of the
particles. At low concentration, this pressure is proportional to the
concentration of particles in contact with a flat surface $%
p_{osm}=k_{B}T\rho _{\infty }$. Indeed, we see from Eq.~(\ref{pression}) that integration over the colloid
positions leads to an effective Hamiltonian for the membrane $\mathcal{H}_{eff}=\mathcal{H}_0+p_{osm}
\int d^2 \mathbf{\rho} h(\mathbf{\rho} )$. The
average position of the membrane is then shifted to its new value $%
\left\langle h\right\rangle =-z_{0}$, so that the concentration profile (or
the dividing surface) is translated from the same distance. For a symmetric
system (particles on both sides with the same chemical potential), this
length would just vanish. Note that when $z=-z_{0}$, the local particle
density is reduced to half of the bulk value, that is $\rho _{\infty }/2$.

Now, let us compute the \emph{excess} particle density, defined as the first
moment of the density profile~\cite{rowlinsonbook} 
\begin{equation}
\Gamma =\int_{-\infty }^{+\infty }dz\left[ \rho \left( z\right) -\rho
_{\infty }\times \theta (z+z_0)\right] \ .  \label{defadsorption}
\end{equation}
$\Gamma $ is called the \emph{adsorption} of the species, since a large
positive value of $\Gamma $\ is an evidence for particle accumulation at the
surface. Conversely, a negative value of $\Gamma $ indicates that
concentration in the surface vicinity is lower than concentration in the
bulk phase. Using relation~(\ref{density}), we find after a simple
integration 
\begin{equation}
\Gamma =0 \ ,  \label{adsorption}
\end{equation}
meaning that there is, on average, neither accumulation nor depletion of
particles near the membrane. Actually, this result comes from the
cancellation of two effects: insertion of particles into the holes and
valleys of the rough surface exactly compensates for the depletion from the
convex regions, as can be seen in fig.~(\ref{fig1}). Eq.~(\ref{adsorption})
also implies that there is no additional contribution to the interfacial
tension, and consequently no spontaneous curvature of the membrane induced
by the colloids. However, this result is only valid for vanishing particle
radius $a $. Although finite size effects are not easily included in the
theory, one does not expect these results to hold for $a\sim \xi _{\parallel
}$ any longer~\cite{lipowskyEPL98,bickelJCP03}.

\section{Radial distribution function and effective potential}

\label{sec:pair}

To better characterize the statistical properties of the system, we now
focus on the pair distribution function defined as 
\begin{equation}
g\left( \mathbf{r},\mathbf{r}^{\prime }\right) =\frac{\sum_{i\neq
j}\left\langle \delta _{3}\left( \mathbf{r}-\mathbf{r}_{i}\right) \delta
_{3}\left( \mathbf{r}^{\prime }-\mathbf{r}_{j}\right) \right\rangle }{\rho
\left( \mathbf{r}\right) \rho \left( \mathbf{r}^{\prime }\right) } \ .
\label{defpair}
\end{equation}
Using the same techniques as before, we find that $g\left( \mathbf{r},%
\mathbf{r}^{\prime }\right) $ can be expressed in terms of the $\varphi _{N}$
functions defined in~(\ref{phin}). Without further detail, we obtain the
formal expression 
\begin{equation}
g\left( \mathbf{r},\mathbf{r}^{\prime }\right) =\frac{\varphi _{2}\left( {%
\boldsymbol{\rho} },{\boldsymbol{\rho} }^{\prime };z+z_{0},z^{\prime
}+z_{0}\right) }{\varphi _{1}\left( {\boldsymbol{\rho} };z+z_{0}\right)
\varphi _{1}\left( {\boldsymbol{\rho} }^{\prime };z^{\prime }+z_{0}\right) }
\ .  \label{pairphin}
\end{equation}
As one could expect, only $\varphi _{i}$'s with $N=1$ and $2$ come out of
the calculations. For the sake of completeness, we give explicitly the
radial distribution function 
\begin{equation}
g\left( \mathbf{r},\mathbf{r}^{\prime }\right) =\frac{\left( 2\pi \right)
^{-1}\left[ \det \mathcal{G}_2\right] ^{-1/2} \int_{-\infty
}^{z+z_{0}}dz_{1}\int_{-\infty }^{z^{\prime }+z_{0}}dz_{2}\exp \left\{ -%
\frac{1}{2}\left( z_{1},z_{2}\right) \mathcal{G}_2^{-1}\binom{z_{1}}{z_{2}}%
\right\} }{\frac{1}{4}\left[ 1+\text{erf} \left( \frac{z+z_{0}}{\sqrt{2}\xi
_{\perp }}\right) \right] \left[ 1+\text{erf} \left( \frac{z^{\prime }+z_{0}%
}{\sqrt{2}\xi _{\perp }}\right) \right] }\ ,  \label{pair}
\end{equation}
where $\mathcal{G}_2$ is the $2\times 2$ correlation matrix 
\begin{equation}
\mathcal{G}_2=\left[ 
\begin{array}{ll}
G\left( 0\right) & G\left( \left| {\boldsymbol{\rho}} - {\boldsymbol \rho }%
^{\prime }\right| \right) \\ 
G\left( \left| {\boldsymbol{\rho}} - {\boldsymbol \rho }^{\prime }\right|
\right) & G\left( 0\right)
\end{array}
\right] \ ,
\end{equation}
and $\det \mathcal{G}_2=G\left( 0\right) ^{2}-$ $G\left(\left| %
{\boldsymbol{\rho}} - {\boldsymbol \rho }^{\prime }\right| \right) ^{2}$. Once
again, the result depends on the relative transverse distance $\left| %
{\boldsymbol{\rho}} - {\boldsymbol \rho }^{\prime }\right| $, whereas it
varies with $z$ and $z^{\prime }$ separately. We emphasize that this pair
correlation function would be identically unity for an ideal gas confined by
a rigid wall. Here, \emph{surface fluctuations} give rise to \emph{bulk
correlations} between colloids over distances that depends both on their
separation and on their upright distance from the membrane. For fixed $z$
and $z^{\prime }$, $g\left( \mathbf{r},\mathbf{r}^{\prime }\right) $ is
maximum when ${\boldsymbol{\rho} }={\boldsymbol{\rho} }^{\prime }$ and
decreases to $1$ as the transverse separation increases. The equality $%
g\left( \mathbf{r},\mathbf{r}^{\prime }\right) =1$ (no correlations) is only
achieved either when $z$ or $z^{\prime }$ goes to $+\infty $ at fixed
parallel distance $\left| {\boldsymbol{\rho} }-{\boldsymbol{\rho} }^{\prime
}\right| $, or when $\left| {\boldsymbol{\rho} }-{\boldsymbol{\rho} }%
^{\prime }\right| \rightarrow +\infty $ at fixed $z$ and $z^{\prime }$.
Indeed, the former requirement expresses that the colloids do not feel the
surface anymore at elevations higher than $\xi _{\perp }$, whereas the
latter asserts that correlations vanish at parallel separation much larger
than the membrane correlation length $\xi _{\parallel }$ (see Appendix~$A$).

With the help of the computed one-- and two--points distribution functions,
we can extract the membrane-induced interactions between particles.
According to Eq.~(\ref{grandcan0}), an effective $N$--body potential $ \mathcal{U}
(\mathbf{r}_{1},..,\mathbf{r}_{N})$ may be defined through 
\begin{equation}
\varphi _{N}\left( {\boldsymbol{\rho} }_{1},..,{\boldsymbol{\rho} }%
_{N};z_{1},..,z_{N}\right) = e^{-\beta \mathcal{U}(\mathbf{r}_{1},..,\mathbf{r}_{N})}
\ .  \label{defpotential}
\end{equation}
The analyze developped in this report indicates that the many-body
interaction decomposes as 
\begin{equation}
\mathcal{U}(\mathbf{r}_{1},...,\mathbf{r}_{N})=\sum_{i}\mathcal{U}_{1}(\mathbf{r}%
_{i})+\sum_{\{i,j\}}\mathcal{U}_{2}(\mathbf{r}_{i},\mathbf{r}_{j})+ \ldots
+  \mathcal{U}_{N} (\mathbf{r}_{1},..,\mathbf{r}_{N}) \ .
\end{equation}
Here, $\mathcal{U}_{1}$ is the effective external potential resulting from thermal
undulations of the membrane. We easily find 
\begin{equation}
\mathcal{U}_{1}\left( \mathbf{r}\right) =-k_{B}T\ln \frac{\rho \left( z\right) }{\rho
_{\infty }} \ .  \label{onebody}
\end{equation}
As the reduced density $\rho \left( z\right) /\rho _{\infty }$ ranges from $%
0 $ to $1$, this potential is always repulsive and tends to move the
particles away from the surface. Note that the corresponding force $%
F=-d\mathcal{U}_{1}/dz$ exerted on a particle by surface undulations remains finite at
``contact'': $F(-z_{0})=k_{B}T/\xi _{\perp }$ (recall that $-z_{0}$ is the
average position of the membrane under the osmotic pressure of the
colloids). Surprisingly, this force increases as the roughness of the
surface $\xi _{\perp }$ decreases, but it has to be this way as $F$
eventually diverges in the hard wall limit.

\begin{figure}[tbp]
\begin{center}
\includegraphics{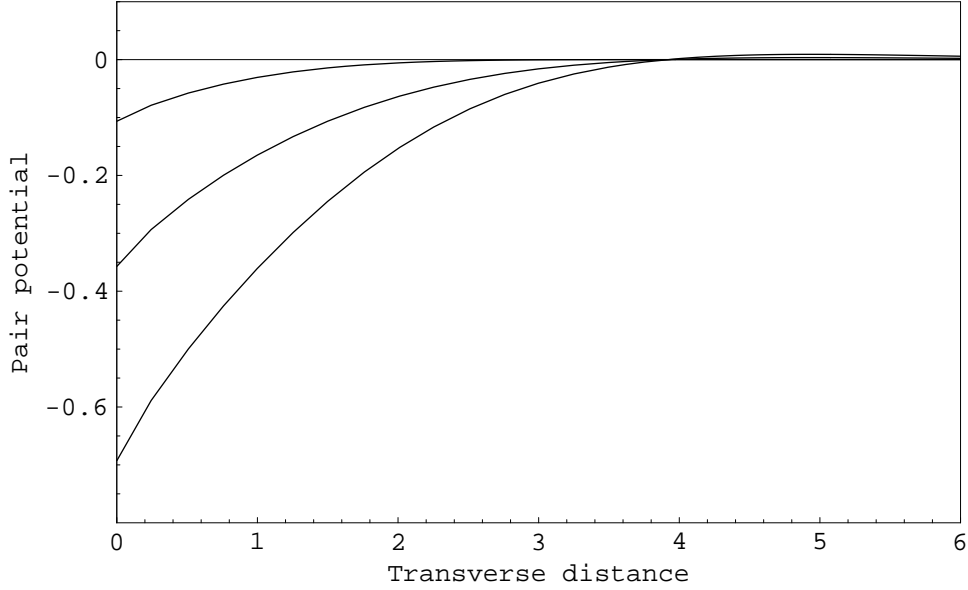} 
\end{center}
\par
\caption{Equal-height, effective pair-potential $\protect\beta \mathcal{U}_{2}(
{\boldsymbol{\rho}} ,{\boldsymbol{\rho}} ^{\prime },z=z^{\prime })$ as a function of the reduced
transverse distance $\left| {\boldsymbol{\rho}} -{\boldsymbol{\rho}} ^{\prime }\right| /%
\xi _{\parallel}$. The curves corresponds to different values of the
perpendicular distance. From bottom to top: $(z+z_{0})/\xi
_{\perp }=0$, $0.5$, and $1$. The depth of the potential well is
of order $k_{B}T$ for $z+z_{0}=0$, and vanishes exponentially at large
separations.}
\label{fig2}
\end{figure}

Regarding the potential of mean force $\mathcal{U}_{2}\left( \mathbf{r},\mathbf{r}%
^{\prime }\right) $, we find 
\begin{equation}
\mathcal{U}_{2}\left( \mathbf{r},\mathbf{r}^{\prime }\right) =-k_{B}T\ln g\left(
\left| {\boldsymbol{\rho}}-{\boldsymbol{\rho}}^{\prime }\right| ,z,z^{\prime
}\right) \ .  \label{twobodies}
\end{equation}
The normalization of $g\left( \mathbf{r},\mathbf{r}^{\prime }\right) $, Eq.~(%
\ref{pairphin}), ensures that only two--body terms are accounted for. The
pair potential $\mathcal{U}_{2}$ is shown in fig.~(\ref{fig2}) for a membrane with no
surface tension ($\sigma =0$) and for fixed $z=z^{\prime }$: it is \emph{%
negative} at short parallel distances and vanishes at large separations.
Accordingly, colloids that are close to the membrane tend to aggregate even
if there are only hard-core repulsions in our description. We can evaluate
the depth of this potential: diagonalizing the quadratic form in~(\ref{pair}%
) leads to 
\begin{equation}
\mathcal{U}_{2}\left( \mathbf{r}=\mathbf{r}^{\prime }\right) =-k_{B}T\ln 2+k_{B}T\ln %
\left[ 1+\text{erf}\left( \frac{z+z_{0}}{\sqrt{2}\xi _{\perp }}\right) %
\right] \ .
\end{equation}
Of course, the interaction still depends on the $z$-position of the pair of
colloids. The depth of the potential increases with decreasing altitude, and
is of order $k_{B}T$ for $z=-z_{0}$. At larger separation, $\mathcal{U}_{2}$ displays
a tiny repulsive barrier ($0.01k_{B}T$ at most). In the limit $d=\left| {%
\boldsymbol{\rho}}-{\boldsymbol{\rho}}^{\prime }\right| \gg \xi _{\parallel
} $, we show in Appendix~$C$ that the potential of mean force vanishes
exponentially 
\begin{equation}
\mathcal{U}_{2}\left( d\right) \sim k_{B}T\sqrt{\frac{\xi _{\parallel }}{d}}e^{-d/\xi
_{\parallel }}\ ,  \label{asymp}
\end{equation}
for $z=z^{\prime }=z_{0}$.

\section{Discussion}

\label{sec:concl}

In this paper, we have adapted the usual many-body statistical mechanics in
order to include an additional degree of freedom, namely the thermal
fluctuations of the membrane. It has been shown that surface undulations
broaden the density profile and generates correlations among an essentially
ideal gas system. As can be seen in Eq.~(\ref{pression}), partial
integration over the positions induces a linear coupling with the membrane
height in the case of point-like particles. For finite-size objects, one
would certainly expect further couplings with membrane curvature as well as
with higher-order terms, but this point is far beyond the scope of this
paper.

As a consequence of surface fluctuations, the colloids attract each other
through the potential of mean force $\mathcal{U}_{2}$. Interstingly, the only
remaining signature of the elastic parameters of the membrane are the length
scales $\xi _{\parallel }$, $\xi _{\perp }$ and $z_{0}$. For two particles at a given position, 
the configurational entropy of the membrane increases as the colloids come closer. This effective pair
interaction is of order of the thermal energy for
particles near the average position of the membrane, and is in many respects similar to the familiar depletion
interaction. In particular, it would simply sum up with a direct
colloid-colloid potential in a more realistic system. Note that our approach is to
some extent peculiar, in the sense that we trace out the degrees of freedom
of the ``slow'' variable, ending up with an effective Hamiltonian for the
small particles. This procedure usually leads to a very poor description of
the system, because one has generally to resort to uncontrolled
approximations. Here, the situation is more refined as we managed to perform
exact calculations. The potential of mean force~(\ref{twobodies}) is
therefore expected to be very accurate for colloids much smaller than the
correlation length $\xi _{\parallel }$.

Finally, we would like to emphasize the fact that $\mathcal{U}_{1}$ and $\mathcal{U}_{2}$ are
the dominant interactions at low concentrations. The framework
developed in this paper would in principle allow us to evaluate the relative
weight of many-body contributions, but these would only be relevant at the
onset of an hypothetical aggregation. Here however, the drift force from the
membrane always move the particle away from the surface so that two--body
attraction is not prevailing. One could still imagine to enforce particle
accumulation near the membrane through a small attractive colloid-membrane
interaction. Whether membrane fluctuations could then induce surface
cristallization is an interesting point, work on this question is currently
under progress.

\acknowledgments

The authors are much indebted to Professors C.\ Misbah and G. Porte for
illuminating discussions during the \textit{Spring School on Soft-Matter
Physics and its Applications}, 12-16 April 2004, Marrakech, Morocco.

\appendix

\section{}

\label{app:a}

In this appendix, we recall a couple of results on fluid membranes. The
height-height correlation function is defined as 
\begin{eqnarray}
G\left( {\boldsymbol{\rho} }\right) &=&\left\langle h\left( {%
\boldsymbol{\rho} }\right) h\left( 0\right) \right\rangle _{0}-\left\langle
h\left( {\boldsymbol{\rho} }\right) \right\rangle _{0}\left\langle h\left(
0\right) \right\rangle _{0}  \nonumber \\
&=&\int \frac{d^{2}\mathbf{q}}{\left( 2\pi \right) ^{2}}\frac{e^{i\mathbf{q}.%
{\boldsymbol{\rho} }}}{\widehat{\kappa } q^{4}+\widehat{\sigma }q^{2}+%
\widehat{\mu }} \ ,
\end{eqnarray}
with the notation $q=\left| \mathbf{q}\right| $. Here, the thermal averages $%
\left\langle \ldots \right\rangle _{0}$ are performed with the Helfrich
Hamiltonian~(\ref{helfrich}) in the absence of particle. For a bilayer
without surface tension $\left( \sigma =0\right) $, the integral over the
Fourier modes leads to 
\begin{equation}
G\left( \rho \right) =-\frac{4}{\pi }\xi _{\perp }^{2}\text{kei}\left( \sqrt{%
2}\frac{\rho }{\xi _{\parallel }}\right) \ ,  \label{correlationmembrane}
\end{equation}
with kei$\left( x\right)= \text{Im} \left[ K_{0}\left( xe^{i\pi /4}\right) %
\right] $ is a Kelvin function~\cite{abrabook}. We then define the mean roughness of the
membrane $\xi _{\perp }=G\left( 0\right) ^{1/2}=2^{-3/2}(\widehat{\kappa }%
.\widehat{\mu })^{-1/4}$, and the in-plane correlation length $\xi _{\parallel
}=2^{1/2}\left( \widehat{\kappa }/\widehat{\mu }\right) ^{1/4}$
characterizing the exponential decay of $G\left( \rho \right) $ at large
distances 
\begin{equation}
G\left( \rho \right) \sim e^{-\rho /\xi _{\parallel }}\qquad ,\qquad \rho
>>\xi _{\parallel } \ .
\end{equation}

\section{}

\label{app:b}

The aim of this appendix is the proof of formula~(\ref{phin}) that defines
the function $\varphi _{N}$. To this end, we first compute its multiple
derivative 
\begin{equation}
\frac{\partial ^{N}\varphi _{N}}{\partial z_{1}...\partial z_{N}}%
=\mathcal{Z}_{0}^{-1}\times \int \mathcal{D}h e^{-\beta \mathcal{H}\left[ h\right]
}\prod_{j=1}^{N}\delta \left[ z_{j}-h\left( x_{j},y_{j}\right) \right] \ .
\label{multiple}
\end{equation}
Write the integral form of the Dirac distribution 
\begin{equation}
\delta \left[ z_{j}-h\left( x_{j},y_{j}\right) \right] =\int_{-\infty
}^{+\infty }\frac{d k_{j}}{2\pi }e^{ik_{j}.\left[ z_{j}-h\left(
x_{j},y_{j}\right) \right] } \ ,
\end{equation}
allows us to rewrite Eq.~(\ref{multiple}) as 
\begin{equation}
\frac{\partial ^{N}\varphi _{N}}{\partial z_{1}...\partial z_{N}}
=\int_{-\infty }^{+\infty }\prod_{j=1}^{N}\left[ \frac{d k_{j}}{2\pi }
e^{ik_{j}.z_{j}}\right] \times \frac{\int \mathcal{D}he^{-\beta \mathcal{H}%
\left[ h \right] +\int d^{2}{\boldsymbol{\rho} }J\left( {\boldsymbol{\rho} }%
\right) h\left( {\boldsymbol{\rho} } \right) }}{\mathcal{Z}_{0}} \ .  \label{funct}
\end{equation}
Here, we have introduced the source 
\begin{equation}
J\left( {\boldsymbol{\rho} }\right) =-i\sum_{j=1}^{N}k_{j}\delta _{2}\left( {%
\boldsymbol{\rho} }-{\boldsymbol{\rho} }_{j}\right) \ ,  \label{defsource}
\end{equation}
with the two-dimensional vector ${\boldsymbol{\rho} }_{j}=\left(
x_{j},y_{j}\right) $. The functional integration in relation~(\ref{funct})
is trivial, since the Hamiltonian $\mathcal{H}\left[ h\right] $ is quadratic
in the field $h$. We simply give the result 
\begin{equation}
\frac{\int \mathcal{D}he^{-\beta \mathcal{H}\left[ h\right] +\int d^{2}{%
\boldsymbol{\rho} } J\left( {\boldsymbol{\rho} }\right) h\left( {%
\boldsymbol{\rho} }\right) }}{Z_{0}}=\exp \left\{ \frac{1}{2}\int d^{2}{%
\boldsymbol{\rho} }\int d^{2}{\boldsymbol{\rho} }^{\prime }J\left( {%
\boldsymbol{\rho}}\right) G\left( {\boldsymbol{\rho} }-{\boldsymbol{\rho} }%
^{\prime }\right) J\left( {\boldsymbol{\rho}}^{\prime }\right) \right\} \ ,
\label{result}
\end{equation}
where $G\left( {\boldsymbol{\rho} }-{\boldsymbol{\rho} }^{\prime }\right) $
is the membrane propagator, relation (\ref{defpropagator}). Now, replace the
source $J\left( {\boldsymbol{\rho} }\right) $ by its definition~(\ref
{defsource}) to find 
\begin{equation}
\frac{1}{2}\int d^{2}{\boldsymbol{\rho} }\int d^{2}{\boldsymbol{\rho} }%
^{\prime }J \left( {\boldsymbol{\rho} }\right) G\left( {\boldsymbol{\rho} }-{%
\boldsymbol{\rho} }^{\prime }\right) J \left( {\boldsymbol{\rho} }^{\prime
}\right) =-\frac{1}{2}\sum_{j=1}^{N}k_{i}G\left( {\boldsymbol{\rho} } _{i}-{%
\boldsymbol{\rho} }_{j}\right) k_{j} \ .
\end{equation}
We therefore obtain 
\begin{eqnarray}
\frac{\partial ^{N}\varphi _{N}}{\partial z_{1}...\partial z_{N}}
&=&\int_{-\infty }^{+\infty }\prod_{j=1}^{N}\frac{dk_{j}}{2\pi }
e^{ik_{j}.z_{j}}\exp \left\{ -\frac{1}{2}\sum_{i,,j=1}^{N}k_{i}.[\mathcal{G}%
_N] _{ij}.k_{j}\right\}  \nonumber \\
&=&\left( 2\pi \right) ^{-N/2}\left[ \det \mathcal{G}_N\right] ^{-1/2}\times
\exp \left\{ -\frac{1}{2}\sum_{i,j=1}^{N}z_{i}^{\prime }.\left[ \mathcal{G}%
_N^{-1} \right] _{ij}.z_{j}^{\prime }\right\} \ .
\end{eqnarray}
The $N^{2}$ coefficients $[\mathcal{G}_N]_{ij}=G\left( {\boldsymbol{\rho} }%
_{i}-{\boldsymbol{\rho} }_{j}\right) $, $1\leq i,j\leq N$, define a squared
matrix $\mathcal{G}_N$. Remark that the above multiple integral is Gaussian.
A straightforward integration yields the explicit expression of the function 
$\varphi _{N}$, formula~(\ref{phin}).

\section{}
\label{app:c}

At large transverse separation $d=\left| {\boldsymbol{\rho} }-{%
\boldsymbol{\rho} }^{\prime }\right| \gg \xi _{\parallel }$, it is possible
to evaluate completely the two-points function $\varphi _{2}$. Indeed, the
propagator satisfies in this limit $G(d)\ll G(0)=\xi _{\perp }^{2}$, so that
we are naturally lead to define the small parameter $\alpha =G(d)/G(0)$. At
first order, one  has $\det \mathcal{G}=\xi _{\perp }^{2}(1+\mathcal{O}%
(\alpha ^{2}))$, and 
\begin{equation}
\left( z_{1},z_{2}\right) \mathcal{G}^{-1}\binom{z_{1}}{z_{2}}=\frac{%
z_{1}^{2}}{\xi _{\perp }^{2}}+\frac{z_{2}^{2}}{\xi _{\perp }^{2}}-2\alpha 
\frac{z_{1}z_{2}}{\xi _{\perp }^{2}}+\mathcal{O}(\alpha ^{2}) \ .
\end{equation}
Expanding the exponential in Eq.~(\ref{phin}) up to first order, we
find 
\begin{eqnarray}
\varphi _{2}(d,z,z' ) &=&\left( 2\pi \right) ^{-1}\xi _{\perp }^{-2}
\int_{-\infty }^{z+z_{0}}dz_{1}\int_{-\infty }^{z^{\prime
}+z_{0}}dz_{2}\exp \left\{ -\frac{z_{1}^{2}+z_{2}^{2} }{2\xi _{\perp }^{2}}%
\right\} \left( 1+\alpha \frac{z_{1}z_{2}}{\xi _{\perp }^{2}}+\mathcal{O}%
(\alpha ^{2})\right)  \nonumber \\
&=&\varphi _{1}(z)\varphi _{1}(z' )+\frac{G(d)}{2\pi \xi _{\perp }^{2}}
\exp \left\{ -\frac{(z+z_{0})^{2}}{2\xi _{\perp }^{2}}\right\} \exp \left\{
- \frac{(z^{\prime }+z_{0})^{2}}{2\xi _{\perp }^{2}}\right\} +\mathcal{O}%
(\alpha ^{2}) \ .
\end{eqnarray}
Taking the logarithm of this expression and using the definition~(\ref
{correlationmembrane}) for $G(d)$ finally leads to Eq.~(\ref{asymp}).

\newpage

\end{document}